# Anisotropic phase diagram and strong coupling effects in $Ba_{1-x}K_xFe_2As_2$

U. Welp[1], R. Xie[1,2], A. E. Koshelev[1], W. K. Kwok[1]

[1]Materials Science Division, Argonne National Laboratory, Argonne, IL 60439

[2]Physics Department, University of Notre Dame, IN 46556, USA

H. Q. Luo, Z. S. Wang, G. Mu and H. H. Wen

Institute of Physics, Chinese Academy of Sciences, Beijing 100190, China

We present a thermodynamic study of the phase diagram of single-crystal $Ba_{1-x}K_xFe_2As_2$ using specific heat measurements. In zero-magnetic field a clear step in the heat capacity of $\Delta C/T_c = 0.1$ J/f.u.$K^2$ is observed at $T_c \approx 34.6 K$ for a sample with x = 0.4. This material is characterized by extraordinarily high slopes of the upper critical field of $\mu_0 \partial H_{c2}^c/\partial T = -6.5$ T/K and $\mu_0 \partial H_{c2}^{ab}/\partial T = -17.4$ T/K and a surprisingly low anisotropy of $\Gamma \sim 2.6$ near $T_c$. A consequence of the large field scale is the effective suppression of superconducting fluctuations. Using thermodynamic relations we determine Ginzburg-Landau parameters of $\kappa_c \sim 100$ and $\kappa_{ab} \sim 260$ identifying $Ba_{1-x}K_xFe_2As_2$ as extreme type-II. The large value of the normalized discontinuity of the slopes of the specific heat at $T_c$, $(T_c/\Delta C)\, \Delta(dC/dT)_{T_c} \sim 6$ indicates strong coupling effects in $Ba_{1-x}K_xFe_2As_2$.

The recent discovery of superconductivity in LaFeAsO$_{1-x}$F$_x$ [1] at 26 K has led to the emergence of two new families of layered high-temperature superconductors: the electron-doped 1111-compunds with compositions REFeAsO$_{1-x}$F$_x$ and rare earths RE = Sm, Ce, Nd, Pr, Gd, Tb, Dy, and the hole-doped 122-compounds [2] with composition A$_{1-x}$K$_x$Fe$_2$As$_2$ and A = Ba, Sr. The substitution of Fe with Co or Ni leads to electron-doped 122-materials [3,4]. Currently, SmFeAsO$_{1-x}$F$_x$ has the highest value of T$_c$ of the 1111-compunds at 55 K [5] and Ba$_{1-x}$K$_x$Fe$_2$As$_2$ (x~0.4) has the highest T$_c$ of the 122-compounds at 38 K [2]. Common to both groups is that the parent compounds undergo a transition to an antiferromagnetic ground state that is accompanied by a structural transition [6]. Upon doping, these transitions are suppressed and superconductivity arises. This proximity to magnetism might lead to unconventional superconducting pairing and gaps. The availability of sizable single crystals of the 122-compounds [7,8] has enabled measurements that have greatly advanced the understanding of Ba$_{1-x}$K$_x$Fe$_2$As$_2$ for instance through the determination of the superconducting gap structure and Fermi surface in ARPES experiments [9-14]. The upper critical field, H$_{c2}$, and its anisotropy are fundamental bulk characteristics that shed additional light on the microscopic length scales and the Fermi surface topology of this new superconductor.

Here we present a thermodynamic study of the phase diagram of single-crystal Ba$_{1-x}$K$_x$Fe$_2$As$_2$ using specific heat measurements. In zero-magnetic field a clear step in the heat capacity of $\Delta C/T_c$ = 0.1 J/f.u.K$^2$ is observed at $T_c \approx 34.6 K$. Using an entropy conserving construction we determine the field dependence of the transition temperature and observe extraordinarily high slopes of the upper critical field $\mu_0 \partial H_{c2}^c / \partial T$ =-6.5 T/K and $\mu_0 \partial H_{c2}^{ab} / \partial T$ =-17.4 T/K and a surprisingly low anisotropy of $\Gamma$ ~ 2.6 near T$_c$. A consequence of this large field scale is the effective suppression of superconducting fluctuations. A thermodynamic analysis yields Ginzburg-Landau parameters of $\kappa_c \sim 100$ and $\kappa_{ab} \sim 260$ identifying $Ba_{1-x}K_xFe_2As_2$ as extreme type-II. The large value of the normalized discontinuity of the slopes of the specific heat at $T_c$, $(T_c/\Delta C)\,\Delta(dC/dT)_{T_c} \sim 6$ indicates strong coupling effects in $Ba_{1-x}K_xFe_2As_2$.

The $Ba_{1-x}K_xFe_2As_2$ crystals used in this study have a nominal composition of x = 0.4 and were grown from a FeAs-self flux as described in [8]. We performed the calorimetric measurements using a membrane-based steady-state ac-micro-calorimeter [15]. It utilizes a thermocouple composed of Au-1.7%Co and Cu films deposited onto a 150 nm thick $Si_2N_4$-membrane as thermometer. A $Ba_{0.6}K_{0.4}Fe_2As_2$ crystal with approximate size of 155x152x19 $\mu m^3$ was mounted onto the thermocouple using Apiezon N grease (see inset of Fig. 1b). An ac-heater current at 23.2 Hz is adjusted such as to induce oscillations of the sample temperature of 50 to 200 mK.

Figure 1 shows the specific heat as C/T (J/f.u. $K^2$) for various fields applied parallel to the c-axis and parallel to the ab-planes, respectively. In zero-field the specific heat displays a clear step-like anomaly with a width of ~1K at the superconducting transition. With increasing applied field this step broadens and shifts systematically to lower temperatures. The transition width increases roughly like $H^{2/3}$ (inset of Fig. 1b) indicative for field-induced superconducting fluctuation effects (see below). In contrast to the data in Fig. 1a, for the single-layer $NdFeAsO_{0.82}F_{0.18}$ compound a magnetic field of about 3 T along the c-axis almost completely suppresses the specific heat anomaly [16]. The reason for this different behavior lies in superconducting fluctuations and their field dependence. Whereas the zero-field Ginzburg number $G_i = \left(8\pi^2 k_B T_c \lambda_{ab}^2 / \Phi_0^2 \xi_c\right)^2 / 2$ is roughly the same for both materials, $G_i \sim 2 \cdot 10^{-3}$, the in-field Ginzburg numbers [17] $G_i(H) = \left(H/H_{c2}^{GL}(0)\right)^{2/3} G_i^{1/3}$ are strongly different. Here, $H_{c2}^{GL}(0) = -(dH_{c2}/dT)_{T_c} \cdot T_c$ is the zero-temperature GL upper critical field. Since for $Ba_{0.6}K_{0.4}Fe_2As_2$ $H_{c2}^{GL}(0)$ is about 6 times larger than for $NdFeAsO_{0.82}F_{0.18}$ a field of 18 T would be required to achieve the same effect as 3 T induced for $NdFeAsO_{0.82}F_{0.18}$ thus accounting for the reduced effect of superconducting fluctuations in $Ba_{1-x}K_xFe_2As_2$.

We use an entropy conserving construction [16,18,19] to approximate the measured transition by an ideal step as indicated in Fig. 1a thereby determining the transition temperature as function of field. The resulting phase diagram is shown in Fig. 2. For fields below 8 T the upper critical field is linear in temperature with slopes of $\mu_0 \partial H_{c2}^c / \partial T = -6.5$ T/K and $\mu_0 \partial H_{c2}^{ab} / \partial T = -17.4$ T/K corresponding to a low superconducting anisotropy of $\Gamma \sim 2.6$. These values for the upper critical field slopes and anisotropy are larger than determinations based on magneto-resistance measurements on similar single crystals [20]. Considering the low value of the anisotropy these slopes are extraordinarily high. For comparison, $Nb_3Sn$ has upper critical field slope of -2.3 T/K whereas that of the Chevrel phase $PbMo_6S_8$ is -6 to -7 T/K and that of $YBa_2Cu_3O_{7-\delta}$ (anisotropy $\sim$ 5-7) is around -12 T/K for H $\|$ ab. The upper critical field slopes and anisotropy of $Ba_{0.6}K_{0.4}Fe_2As_2$ are though comparable to those of the heavy Fermion superconductor $CeCoIn_5$ with $T_c \sim 2.4$ K [21]. The zero-temperature Ginzburg-Landau coherence lengths of $Ba_{0.6}K_{0.4}Fe_2As_2$ are $\xi_{ab} \approx 1.2nm$ and $\xi_c \approx 0.45nm$, and the zero-temperature upper critical fields can be estimated using the single-band Werthamer-Helfand-Hohenberg (WHH) formula $H_{c2}(0) = -0.69 \ T_c \ (\partial H_{c2}/\partial T)_{T_c}$ as $H_{c2}^c(0) \approx 155 \ T$ and $H_{c2}^{ab}(0) \approx 415 \ T$. These estimates are clearly larger than the paramagnetic limiting field $\mu_0H_P[T] = 1.84\, T_c[K] = 64$ T for weak spin-orbit scattering. The Maki parameter $\alpha = \sqrt{2}H_{c2}(0)/H_P$ describing the relative strength of orbital pair breaking and paramagnetic limiting, reaches values of 3.4 and 9 for the c-axis and ab-plane, respectively. Since $Ba_{1-x}K_xFe_2As_2$ is composed mostly of low-Z elements spin-orbit scattering may be weak and paramagnetic limiting effects such as a first order transition into the normal state might arise at low temperatures as have been observed for $CeCoIn_5$ [21,22].

Recently, high-field measurements of the magneto-resistance [23] and rf-penetration [24] on crystals of $Ba_{1-x}K_xFe_2As_2$ with critical temperatures of 28 K and 29.5 K, respectively, and of the magneto-resistance [25] of a $Ba(Fe_{0.9}Co_{0.1})_2As_2$ crystal with $T_c \sim 22$ K have been performed. The observed upper critical fields lie well below 100 T in accordance with the trend that $H_{c2}(0)$ is strongly reduced for decreasing $T_c$ [20] thereby strongly decreasing the Maki parameter as well. A remarkable finding in these high-field measurements is the strong decrease of the anisotropy with decreasing temperature. This behavior is in contrast to that of $SmFeAsO_{0.8}F_{0.2}$ [26] and the two-band superconductor $MgB_2$ [27] for both of which the anisotropy increases with decreasing temperature. In the temperature range close to $T_c$ covered in Fig. 2 our data reveal – within the experimental uncertainties – a temperature independent anisotropy.

Band structure calculations [10,28] of the hole doped $Ba_{1-x}K_xFe_2As_2$ reveal four Fermi surface sheets, two concentric corrugated cylindrical electron surfaces around the X-point of the body centered tetragonal Brillouin zone and two hole barrels around the Γ-point. The outer hole barrel is characterized by pronounced z-axis dispersion giving it quasi 3D character. This dispersion is found to depend sensitively on the doping level and the position of the As-atoms with respect to the Fe-sheet. The hole surfaces have low and the electron cylinders have high Fermi velocities. ARPES measurements [9-14,29] are largely consistent with this structure. The quasi-3D band could account for the low anisotropy of $Ba_{1-x}K_xFe_2As_2$. However, ARPES measurements also reveal a large superconducting gap on the inner hole barrel and on the electron cylinders, and a small gap on the outer hole barrel. Therefore, at low temperatures and high fields where the large superconducting gaps on the more 2D Fermi surface sheets become dominant the anisotropy should increase. This is the behavior of $MgB_2$ where the large gap resides on the 2D σ-bands and the smaller gap on the 3D π-bands. The observed decrease of the anisotropy of $Ba_{1-x}K_xFe_2As_2$ with decreasing temperature suggests a different mechanism such as paramagnetic limiting as discussed above.

The entropy conserving construction in Fig. 1a yields a zero-field jump in the specific heat of $\Delta C/T_c = 0.1$ J/f.u. $K^2$, in good agreement with a previous report [30]. This jump in the specific heat – when expressed in units of $J/m^3K^2$ – is related to the thermodynamic critical field $H_c$ through the relation $\mu_0 \Delta C/T_c = (\mu_0 dH_c/dT)^2_{T_c}$ resulting in a slope of -45 mT/K of $\mu_0 H_c$ at $T_c$. Using the single-band weak-coupling BCS-relation the zero-temperature thermodynamic critical field of $\mu_0 H_c(0) = -0.576 \cdot (dH_c/dT)_{T_c} \cdot T_c \approx 0.9 T$ can be estimated. Strong coupling effects would reduce this value. In the Ginzburg-Landau regime, $H_c(T)$ is linear in temperature and is given by $H_c(T) = \Phi_0/(2\sqrt{2}\pi\mu_0 \lambda^{GL}_{ab}(T)\xi_{ab}(T))$ with $\lambda^{GL}_{ab}(T) = \lambda^{GL}_{ab}/(1-T/T_c)^{1/2}$ which yields an in-plane GL-penetration depth of $\lambda^{GL}_{ab} \approx 125 nm$. Then, the Ginzburg-Landau parameters $\kappa_c = \lambda^{GL}_{ab}/\xi_{ab} \approx 100$ and $\kappa_{ab} = (\lambda^{GL}_{ab}\lambda^{GL}_c)^{1/2}/(\xi_{ab}\xi_c)^{1/2} \approx 260$ can be obtained identifying this material as extreme type-II. Accordingly, the slopes of the lower critical fields $dH_{c1}^i/dT \cong dH_c/dT \, (\ln(\kappa_i) + 0.5)/(\sqrt{2}\kappa_i)$ are low: $\mu_0 dH_{c1}^c/dT \approx -1.6 mT/K$ and $\mu_0 dH_{c1}^{ab}/dT \approx -0.74 mT/K$, respectively. Recent magnetic measurements [31] of the lower critical field of similar samples yielded slopes of $H_{c1}^c$ at $T_c$ of ~-1.8 mT/K underlining the consistency of our thermodynamic analysis. The actual low-temperature penetration depth is larger than the GL value and can be estimated [32] as $\lambda_{ab} \approx \sqrt{2}\lambda_{ab}^{GL} \approx 175 nm$ which is in reasonable agreement with experimental determinations from optical measurements [33].

Evidence for strong coupling effects arises from the value of the normalized discontinuity of the slopes of the specific heat at $T_c$, $(T_c/\Delta C) \, \Delta(dC/dT)_{T_c}$. In single-band weak-coupling BCS theory this ratio is 2.64. In strong-coupling Pb it is 4.6 [32] whereas in the two-band superconductor $MgB_2$ a value of 3.35 can be deduced [19]. From the data in Fig. 1a we obtain a very large value of $(T_c/\Delta C) \, \Delta(dC/dT)_{T_c} \sim 6$ indicative of pronounced strong-coupling effects. Strong coupling can also be inferred from the large value of the gap ratio $2\Delta/k_B T_c \sim 6.8 - 8$ [14], which in weak coupling BCS theory is 3.53. We note, though, that in the presence of multiple superconducting gaps ratios such as $\Delta C/\gamma_n T_c$, $2\Delta/k_B T_c$ or $(T_c/\Delta C) \, \Delta(dC/dT)_{T_c}$ are not universal constants but will depend on the specifics of the gaps.

Fig. 3a shows the angular dependence of the transition temperature $T_{c2}(\theta)$ in a field of 6 T. Within the effective mass model of the Ginzburg-Landau theory of anisotropic superconductors, and assuming linear phase boundaries, $T_{c2}(\theta)$ is given by $T_{c2}(\theta) = T_{c0} + H\sqrt{\cos^2(\theta) + \Gamma^2 \sin^2(\theta)}/(\partial H_{c2}^{ab}/\partial T)$. Here, $\theta$ is the angle of the magnetic field with respect to the FeAs-planes and $T_{c0}$ is the zero-field transition temperature. A fit with $\Gamma = 2.56$ describes the data reasonably well. However, there are small systematic deviations by which the measured values fall below the GL-fit. These are more clearly visible when plotting the deviations $\left(T_{c2}^{GL} - T_{c0}\right)^2 / \left(T_{c2} - T_{c0}\right)^2$ as function of $\cos^2(\theta)$, Fig. 3b. A similar trend in the data has been observed for $MgB_2$, which could be accounted for in a model of two-band superconductivity [34]. Although detailed calculations have not yet been performed the data in Fig. 3 are suggestive for multiple superconducting gaps in $Ba_{0.6}K_{0.4}Fe_2As_2$.

In summary, we have determined the upper critical field of single-crystal $Ba_{0.6}K_{0.4}Fe_2As_2$ using specific heat measurements. The upper critical field slopes are very high, $\partial H_{c2}^{c}/\partial T = -6.5$ T/K and $\partial H_{c2}^{ab}/\partial T = -17.4$ T/K, which correspond – in a single-band model – to zero-temperature GL coherence lengths of $\xi_{ab} \approx 1.2 nm$ and $\xi_c \approx 0.45 nm$ and a low superconducting anisotropy of $\Gamma \sim 2.6$. The very high critical field slopes my enable the observation of paramagnetic limiting effects at low temperatures. A thermodynamic analysis yields Ginzburg-Landau parameters of $\kappa_c \sim 100$ and $\kappa_{ab} \sim 260$ identifying $Ba_{1-x}K_xFe_2As_2$ as extreme type-II. The large value of the normalized discontinuity of the slopes of the specific heat at $T_c$, $(T_c/\Delta C)\, \Delta(dC/dT)_{T_c} \sim 6$ indicates strong coupling effects in $Ba_{1-x}K_xFe_2As_2$.

This work was supported by the US Department of Energy – Basic Energy Science – under contract DE-AC02-06CH11357, by the Natural Science Foundation of China, the Ministry of Science and Technology of China (973 project No. 2006CB60100, 2006CB921802, 2006CB921107) and the Chinese Academy of Sciences (Project ITSNEM).

Figure captions

Fig. 1. Temperature dependence of the specific heat of $Ba_{0.6}K_{0.4}Fe_2As_2$ in units of J/f.u. $K^2$ near the superconducting transition in various fields applied parallel to the c-axis (a) and parallel to the FeAs-planes (b). The tetragonal unit cell contains two formula units (f.u.). The solid lines in (a) illustrate the entropy conserving construction used to determine the transition temperature marked by crosses for the 0 T and 8 T data. The dashed line and the arrows indicate the determination of the width of the transition. The inset in (b) shows the transition width in c-axis fields as function of $H^{2/3}$.

Fig. 2. Phase diagram of $Ba_{0.6}K_{0.4}Fe_2As_2$. The slopes of the upper critical field are -6.5 T/K and -17.4 T/K for the c-axis and ab-planes, respectively.

Fig. 3. (a) Angular dependence of the transition temperature $T_{c2}$ in a field of 6 T. The solid line is a fit to the anisotropic effective mass model. (b) The deviations between the experimental data and fit in panel (a) plotted as $\left(T_{c2}^{GL} - T_{c0}\right)^2 / \left(T_{c2} - T_{c0}\right)^2$ versus $\cos^2(\theta)$. The dashed line is a guide to the eye. In this presentation the line at 1 represents the GL-prediction. The inset in (b) shows a photo of the sample mounted onto the calorimeter.

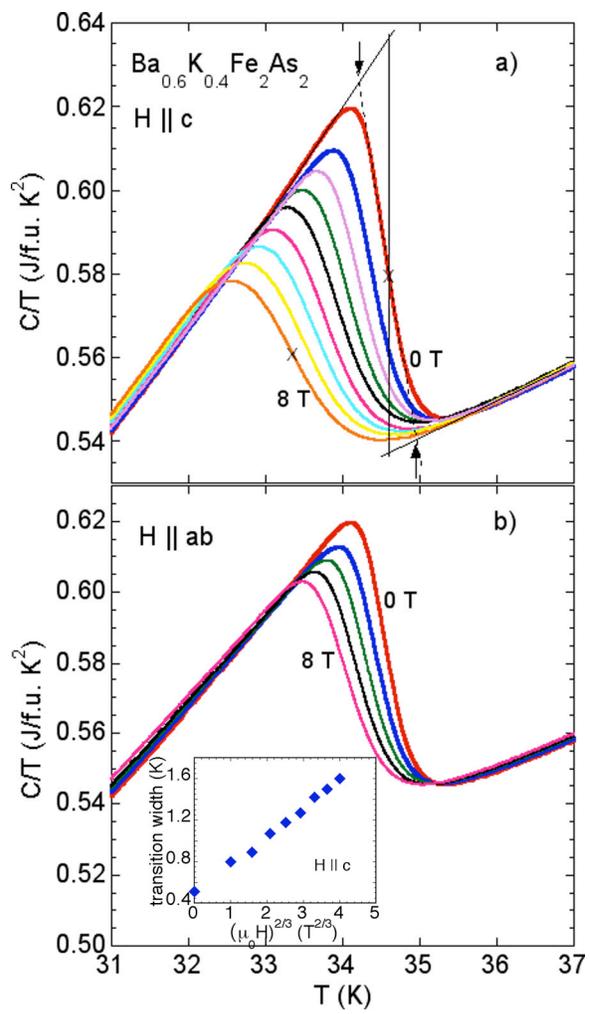

Fig. 1

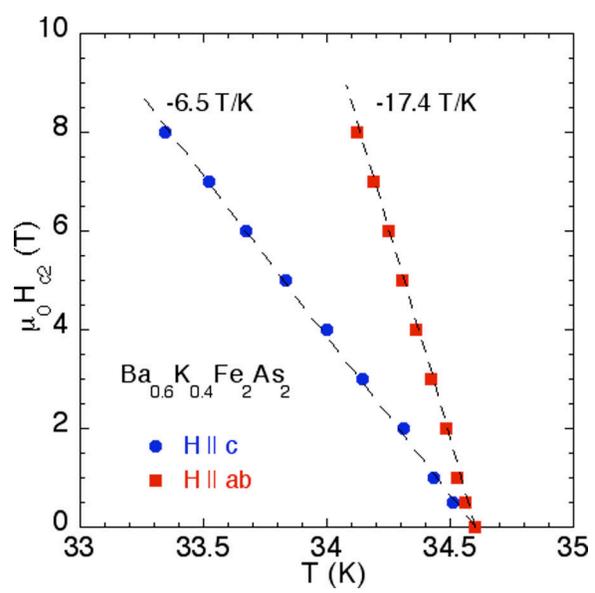

Fig. 2

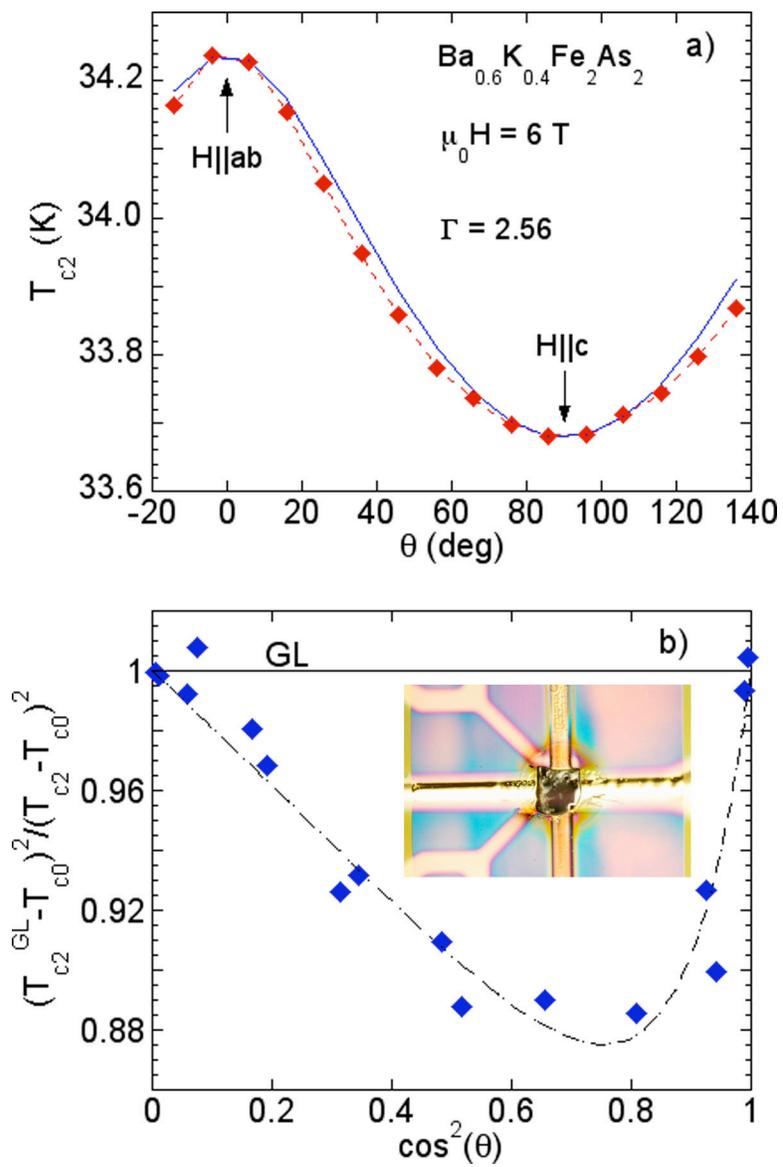

Fig. 3